\documentclass[twocolumn,superscriptaddress,showpacs]{revtex4}
\usepackage{times} 
\usepackage{amsbsy,amssymb,amsmath,bm}
\usepackage{graphicx,color,epsfig,rotate}
\usepackage{fancyhdr}

\def\bbbc{{\mathchoice {\setbox0=\hbox{$\displaystyle\rm C$}\hbox{\hbox
to0pt{\kern0.4\wd0\vrule height0.9\ht0\hss}\box0}}
{\setbox0=\hbox{$\textstyle\rm C$}\hbox{\hbox
to0pt{\kern0.4\wd0\vrule height0.9\ht0\hss}\box0}}
{\setbox0=\hbox{$\scriptstyle\rm C$}\hbox{\hbox
to0pt{\kern0.4\wd0\vrule height0.9\ht0\hss}\box0}}
{\setbox0=\hbox{$\scriptscriptstyle\rm C$}\hbox{\hbox
to0pt{\kern0.4\wd0\vrule height0.9\ht0\hss}\box0}}}}

\newcommand{\ignore}[1]{}
\newcommand{\mComment}[1]{}
\newcommand{\gComment}[1]{}
\newcommand{\jComment}[1]{}
\newcommand{\rComment}[1]{}
\newcommand{\lComment}[1]{}

\renewcommand{\mComment}[1]{\textcolor{blue}{Manny: #1}}
\renewcommand{\gComment}[1]{\textcolor{red}{Gerardo: #1}}
\renewcommand{\jComment}[1]{\textcolor{green}{Jim: #1}}
\renewcommand{\rComment}[1]{\textcolor{magenta}{Ray: #1}}
\renewcommand{\lComment}[1]{\textcolor{purple}{Rolando: #1}}

\begin{document}

\title{The BCS - BEC Crossover In Arbitrary Dimensions}

\author{Zohar Nussinov}
\email{zohar@viking.lanl.gov}
\affiliation{Department of Physics, Washington University, St. Louis, 
MO 63160-4899, USA}
\affiliation{Theoretical Division,
Los Alamos National Laboratory, Los Alamos, NM 87545, USA}

\author{Shmuel Nussinov}
\email{nussinov@post.tau.ac.il}
\affiliation{School of Physics
and Astronomy, Tel Aviv University,
Ramat-Aviv, Tel Aviv 69978, Israel}

\date{Received \today }

\begin{abstract}
 Cold atom traps and certain neutron star layers may contain fermions
 with separation much larger than the range of pair-wise potentials yet
 much shorter than the scattering length.
  Such systems can display {\em universal} characteristics independent of
 the details of the short range interactions.
  In particular, the energy per particle is a fraction $\xi$ of the
 Fermi energy of the free Fermion system.
  Our main result is that for space dimensions D smaller than two and
 larger than four a specific extension of this problem readily yields
 $\xi=1$ for all $D \le 2$ whereas $\xi$ is rigorously 
non-positive (and potentially
vanishing) for all $ D \ge 4$. We discuss the
  $D=3$ case. A particular  unjustified recipe  suggests $\xi=1/2$ in
  $D=3$.
\end{abstract}

\maketitle

\section{Introduction}

Bose-Einstein condensates (BEC) in dilute atomic gases were achieved 
in 1995 for rubidium, sodium, and lithium 
\cite{gas95}. Recently, with the production of cold atomic gases
close to the Feshbach resonance \cite{RGJ}, these systems 
provided a vehicle for the study of the BCS to BEC
crossover. The (BCS) superconducting and 
Fermi superfluid are mixed systems harboring 
quasi-free and pair correlated fermions. In the dilute strong
coupling molecular BEC, all fermions are relatively tightly
bound into pairs forming a unique macroscopic state.
Several theoretical studies have been 
carried, e.g. \cite{DH}. Many of these
works report on very interesting universal aspects of the 
BCS to BEC transition. In such dilute systems with an 
inter-particle separation much greater than 
the range of the pair-wise potential yet far shorter than the scattering
length, the value of the energy density at this crossover 
is independent of specific details. This {\bf universal}
energy density is only dimension dependent.
Direct results on the 
three dimensional system are hard to obtain
and numerical works have been extremely 
valuable. As is well known
from a multitude of other arenas, a dimensional
generalization of original problems posed in three
dimensions to arbitrary dimensions
often allows us an analytical access 
to original three dimensional problems (e.g.
the well known $\epsilon = 4-D$ expansion,
with $D$ the spatial dimension,
which has been extremely fruitful in statistical
and quantum field theories, e.g. \cite{epsilonexpansion}). 
Here, we follow suite and couch the BCS to BEC crossover problem
in arbitrary dimensions ($D$). We report new results 
on the energy per particle
at the onset of this crossover in such an arbitrary dimension
(suitably defined by the appearance of zero-energy two particle bound states). 
This extension enables us to examine crossovers
in low dimensions where no condensed phases occur
due to the enhanced role of low energy fluctuations. 
We will illustrate, both by exact variational bounds and 
normalization considerations, that 
in all dimensions $D \ge 4$ the energy per particle
at the onset of the transition 
is rigorously bounded from above by zero. 
By contrast, due to localization 
tendencies in low dimensions ($ D \le 2$),
each particle carries, on average, the mean
energy of a free Fermi system
at the appearance of the first two 
particle bound states. The physically 
pertinent well known question \cite{ChallProb} 
concerns the energy per particle 
in actual transitions
occurring in $D=3$ dimensions. The determination
of the energy per particle here is far more difficult.
In $D=3$, we depart from the more rigorous 
results in higher and lower dimensions and 
present a heuristic argument for
recently reported numerical results \cite{CCPS}.
These independent heuristic arguments bolster
the the result attained by dimensional interpolation.
The average of the exact bounds on the 
energies in $D=2$ and $D=4$, 
which is half the free fermion energy per particle,
is not far removed from the
numerical results reported in three 
dimensions \cite{CCPS}.
In \cite{CCPS} the fraction of a half, derived here by 
(i) dimensional interpolation and (ii) 
an independent heuristic argument, is 
replaced by $\xi \approx 0.44$.
 
\section{Outline}

We begin, in section(\ref{form}), by formulating
the problem and briefly reviewing numerical results.
Next, in section(\ref{Dim}), we 
turn to an exact variational bound concerning this 
transition in high dimensions and an easier
result relating to bound states in 
low dimensions. 
Finally, in section(\ref{threeD}),
we present heuristic arguments for $D=3$.
These further correlate the 
average energy per particle
with the observed differences 
seen in finite size system 
with an even or odd particle 
number. In a brief appendix,
we provide the specifics of the 
exact solution of the zero energy 
bound state problem in a 
D-dimensional spherical potential
well. Throughout the text, 
in order to make the physics
and scaling very transparent, we will often
use simpler forms. Nevertheless, at the end
of all calculations, we will demonstrate that our
results go unchanged with the insertion
of the exact zero-energy bound state 
solution.

\section{Formulating the Problem: Introductory and General Comments}
\label{form}

We start by writing down the general Hamiltonian. 
describing $N=2n$ spin-1/2 Fermions in a box of size $V=L^3$,
\begin{eqnarray}
  H(g) & = & \sum_{I} \frac{{\bf{p}}_I^2}{2m} - g \sum_{I>J} V(|{\bf{r}}_I-{\bf{r}}_J|) \nonumber
\\     & = & \sum_{i} \frac{{\bf{p}}_i^2}{2m} + \sum_{j} \frac{({\bf{p}}'_j)^2}{2m} \nonumber
\\
& \;\;\;\;\;\;\;\; - & g \sum_{ij} V(|{\bf{r}}_i-{\bf{r}}'_j|).
\label{Hamiltonian} 
\end{eqnarray}
Here, $ I,J $ (or $ i,j $) run from 1 to $N$ (or $n$) respectively.
The phase space coordinates ${\bf r}_i,{\bf p}_i$ and ${\bf r}'_j,{\bf p}'_j$
 are the positions and momenta of the ``spin up" and ``spin down" atoms 
respectively and $[-g V({\bf r}_i,{\bf r}'_j)]$ is the attractive short 
range potential in mixed pairs. These conventions for the 
coordinates [using upper case 
characters to describe the total system (both spin up and spin
down) quantities and the use of lower case characters
(either ``primed'' or ``un-primed'') for spin-down and spin-up
coordinates] will be consistently employed throughout this work.

\begin{figure}
\centerline{\psfig{figure=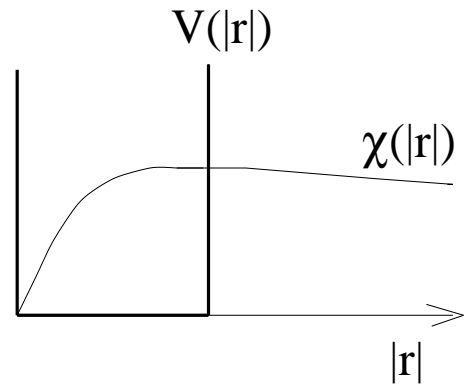,height=
5cm,width=6cm,angle=0}}
\caption{A schematic (thick lines) of the binding spherically symmetric
potential $V({\bf{r}}_{i} - {\bf{r'}}_{j}) \equiv V({\bf{r}}_{ij})$ 
(in three dimensions) which is non-zero only inside a sphere of 
radius $r_{0}$ about the origin, $V({\bf{r}}) = - V_{0} 
\Theta (r_{0} - |{\bf{r}}|)$.
A cartoon of a scaled very weakly bound s-wave state 
($\chi({\bf{r}}_{ij}) = |{\bf{r}}_{ij}| \phi({\bf{r}}_{ij})$)
is shown. To attain exactly one zero energy bound state we
need to fit exactly a quarter period for 
$\chi(|{\bf{r}}|) \sim \sin \kappa r$
and set $ \kappa r_{0} = \frac{\pi}{2}$ with $\kappa 
= \sqrt{2 \mu V_{0}}/\hbar$.
Here, $\mu = m/2$ is the reduced mass of a fermionic pair.
The zero-energy bound state wavefunction in general 
dimension $D$ is derived in the appendix.} 
\label{Fig1bound}
\end{figure}

Atoms of identical spin polarization cannot be 
in a relative S wave state where the short range 
interactions operate and thus, in Eq.(\ref{Hamiltonian}),
we set $V({\bf r}_i,{\bf r}_j)
 = V({\bf r}'_k,{\bf r}'_l)= 0$. The coupling
$g$ is tuned to $g^*$ where the first (zero energy, S wave)
two-body
 bound state appears and the scattering length $a$ diverges. 
An illustration of the standard two-particle zero energy 
bound state in three dimensions 
is provided in Fig.(\ref{Fig1bound}).
With this identification of $g^{*}$ at hand \cite{ChallProb}, 
{\em the BCS to BEC crossover problem is now formally 
defined in arbitrary dimension $D$}, 
although, due to the enhanced role of 
low energy fluctuations, 
actual condensates may form only in $D>2$.
Our physical interest is in $D=3$. 
To make $g$ dimensionless in Eq.(\ref{Hamiltonian},
we scale V by $(m r_0^2)^{-1}$ with $r_0$ the range 
of the potential and $m$ the particle mass. 
We assume a dilute system with inter-particle separation $d=1/k_F$, 
much larger than $r_0$. In practical terms, 
this implies that 
we will always consider the limit $r_{0} \to 0$. 
To avoid the well known formal ``collapse problem'' in which
all particles may sit within the attractive potential
well [as the attractive $V({\bf r}_{i}, {\bf r'}_{j}) <0$ for 
$|{\bf r}_{i} - {\bf r'}_{j}| \le r_{0}$, we may
envision a state in which all particles sit
within a sphere of radius $r_{0}$], 
leading to a divergent negative energy density,
the problem is formulated with the provision that
the limit $r_{0} \to 0$ is taken before the thermodynamic
limit is considered \cite{ChallProb, Baker}. With these
definitions in tow, the problem emulates what transpires 
in many dilute fermionic systems (cold atom traps,
neutron star layers) with separations 
far larger than the range of the pair potentials 
yet far smaller than the scattering length.

In three dimensions,
scaling arguments then imply a universal form for the energy per 
particle
\begin{equation}
                 \frac{E}{N} =  (\frac{3}{5} \, \varepsilon_{F}) \xi,
\label{TotEnergyNbody}  
\end{equation}
with $\varepsilon_F$ the Fermi energy of the free fermi system. \cite{scale}
The fraction $\xi$ is independent of the specific potential
$V({\bf r})$ chosen and is the same for
 all short range potentials which have a zero energy S-wave bound state.
Cold dilute Fermionic atoms display many fascinating universal (and not
universal) features; see, e.g., \cite{DH}. Here we will focus just
on the above $\xi$ parameter.

In actual traps, an excited zero 
energy state $\psi^e_0({\bf r})$ plays a key role. The continuum
 states are effected by the extended zero energy state 
and have negligible overlap with lower, 
tightly bound, states of size $\sim r_0$ which can be 
ignored unless we consider
very long time scales. As any zero 
energy S wave bound state has the same form
 $\psi_0({\bf r})= \psi^e_0({\bf r})= A/|{\bf r}|$
outside the range of the potential
 for $|{\bf r}| > r_0$,  we assume just {\it one} zero energy state. For $g=0$,
there are two free species with $n=N/2$ fermions in each, and
relative to Eq.(\ref{TotEnergyNbody}), with the standard results
\begin{equation}
\frac{E_{free-fermion}}{N}= \frac{3}{5} \varepsilon_F, \;\;\; 
 n= \frac{L^3
(4\pi/3)p_F^3}{(2\pi)^3},
\label{freespecies}        
\end{equation}
where $p_F$ and $ \varepsilon_F= \frac{p_F^2}{2m}$ 
are the free particle Fermi momentum and Fermi energy respectively.
Throughout, we set $\hbar$=1. Thus, by definition, 
\begin{equation}
  \xi=1   \;\;\;    {\rm for} \;\; {\rm g=0}.
\label{xi1}   
\end{equation}
 For small $g>0$ (weak attractive potentials), the BCS wave function is 
adequate. By contrast, for strong coupling
$g|V(|{\bf r}| < r_{0})| >> \varepsilon_{F}$, 
up-down spin pairs tightly bind into dimers
 of sizes of order, $r_0$, the range of the potential V. 
The $n=N/2$ dimers in the deepest bound state then behave 
as point-like bosons and undergo BEC (Bose-Einstein
Condensation) to the traps' ground state with $p_{dimer}\sim 0$ so that
\begin{equation}
               \frac{E}{N} = - \frac{|B.E.|}{2}  \;\;\; {\rm if} \;\; g \rightarrow  \infty,
\label{dimers}   
\end{equation}
with $B.E.$ the binding energy.

The focus of the current work is on 
the crossover between the BCS and BEC regimes
wherein the initial Cooper pairs become tighter dimers
\cite{LeggetEngelbrecht}.
 The existence of a bound state renders 
a weak-coupling perturbation series inappropriate.
Similarly, strong coupling/tight binding schemes also fail. 
As $g$ decreases so does the binding energy. Once $|B.E.| <\varepsilon_F$, the size of the bound state becomes larger
 than the average inter-particle distance: $O(m \varepsilon_F)^{-1/2} \sim d$  and antisymmetrization
 of spin up (and of spin down) atoms---which is negligible for small dimers---raises
 $E/N$ above $\frac{|B.E.|}{2} =0$.

 The difficulty and interest of the problem motivated several calculations.
 A semi-analytic approach utilizing the Pad\'{e} approximation has
been discussed at length in Ref.  \cite{Baker}.  Numerical Fixed
Nodal Planes-Green Function-Monte Carlo (FNPGFMC or MC)
calculations were used
 to estimate $E(N)$\cite{CCPS}. The ground state wave function of $N$ fermions in a
 periodic box of size $ L $ was estimated by evolving 
(in imaginary time $\tau = - it$) an initial trial
 function $\Psi({\bf r}_i,{\bf r'}_j;\tau=0)$ with 
$\exp (-H \tau)$ . The short range potential
 $V({\bf r}_i-{\bf r}'_j)$ was chosen to have precisely one 
zero energy bound state. 
The lowest energy obtained to date via these methods corresponds
to $\xi = 0.44$ in Eq.(\ref{TotEnergyNbody}) which is 
consistent with present imprecise experimental values.
 The $L/d \sim n^{1/3}=80^{1/3} \approx 3.5$ used  
there implies a box size $L$ smaller
 than the
 (infinite) scattering length. This still allows 
for an extraction of reliable data.

  In general, the 
ground state energy $E/N$ decreases continuously with increasing
 $g$. The instability against the formation of Cooper pairs for any
 attractive potential adds a singular $\sim \exp (-g_{c}/g)$ 
(with $g_{c}$ a constant) term to E/N with the continuity in $g $ maintained.

 \section{Extension to arbitrary spatial dimensionality D}
\label{Dim}

  While the original problem is posed in $D=3$ dimensions, it is
instructive to address it in any ({\em continuous}) D with the simple
extension of Eqs.(\ref{TotEnergyNbody},\ref{freespecies}),
\begin{eqnarray}
\frac{E}{N} =\xi \frac{D}{(D+2)} \varepsilon_F.
\label{trivDim}
\end{eqnarray}

To set the ground for future notation, we mention that the simple
$D$ dimensional extension of the latter part of 
Eq.(\ref{freespecies}) has the total fermion 
number (per spin flavor) within a Fermi sphere of radius $p_{F}$ given by
$n= \Omega_{D}L^{D} p_{F}^{D}/(2 \pi)^{D} $. Here,
$\Omega_{D} = \pi^{D/2}/\Gamma(\frac{D}{2} +1)$
is the volume of the D dimensional unit sphere.

In what follows, we 
(i) establish by localization tendencies in dimensions $D \le 2$ 
that $\xi(D \le 2) ={\xi}(g=0)= 1$. We then (ii) illustrate
that $\xi( D \ge 4)  \le  0$ by relying on divergence, at short length
scales, of the wave function normalization. This simple scaling
consideration is then made rigorous for all continuous dimensions
$D >4$ by employing a variational bound with the 
use of an ``orbital'' Slater determinant wavefunction. 
(Regretfully, we have not been
able to devise a wavefunction which will lead to new
stringent variational bounds on the three-dimensional 
problem.)

Putting all of the pieces together, we find by 
extending the problem to any dimension $D$ 
while insisting on having precisely one zero energy bound state 
that $\xi$ is non-trivial only 
within the interval
 $2 <D < 4$. From this perspective, the only interesting integer 
dimensionality is, in fact, the original D=3!

There may be other extensions of the problem which might be non-trivial 
in all dimensions. Many systems simplify at $D \rightarrow \infty$ and 
a mean field approach may apply there also to Fermionic problems \cite{MV}. 
It has been argued that a certain class of diagrams
 dominates in this limit the perturbative series in an appropriate 
effective field theory and can be analytically summed in certain fashions 
to yield ${\xi}(D={\infty}) = \frac{4}{9}$ \cite{Steel} or $\frac{1}{2}$ 
\cite{SKC}.

  In what briefly follows, we employ 
the above continuation with a primary focus on energetics (i.e., the
 existence of a zero energy S wave bound state) rather than 
on the infinite scattering length. We find the resulting simple 
values $\xi = 1$ for all (continuous) dimensions smaller than 
$D=2$ and $\xi \le 0$ for all $D \ge 4$ interesting on their own right.

\subsection{Low Dimensions ($D \le 2$)}

Localization tendencies present in low dimensions afford us with
a direct result. Purely attractive
 potentials $V({\bf r})$ have zero energy bound states in one {\it and} two dimensions
\cite{LL}, for any strength and $g$. It follows that 
a bound state appears immediately as
the two body attraction is introduced, i.e.
$g^* = 0^{+}$. As the energy fraction ${\xi}(g)$ is
continuous in any dimension D, we find that
\begin{equation}
 \xi ={\xi}(g=0)= 1 \; , \;\;  D \le  2.
\label{D2}  
\end{equation}
Similar incarnations of localization tendencies
in low dimensions have proven very useful in 
other arenas (e.g. \cite{gang4} in which
a scaling theory of localization suggested $D=2$ 
as the marginal dimension for the appearance of
metallic states).

\subsection{High Dimensions ($D \ge 4$)}

\subsubsection{Energy attained by Normalization conditions} 
\label{norm}
 
Next, let us assume that the potential V vanishes outside the range $r_0$. 
Volume normalization factors aside, the spin-up spin-down, zero energy,
pair wave function for relative separation
$|{\bf r} |> r_0 $ is $ \phi({\bf r})
= \frac{A_{>}}{|{\bf r}|^{(D-2)}}$, with $A_{>}$ 
a dimension dependent numerical constant.
 The following is
 a very interesting observation: the integral representing the normalization of this
 zero energy wave function, $ [\int d^D {\bf r} |\phi({\bf r})|^2]$ diverges at small
 ${\bf r}$
 for $D \ge 4$. We consequently find that in  
 $D \geq 4$ dimensions most of the normalization
 of the zero
 energy bound state wave function is concentrated near  
$|{\bf r}| = r_{0}^{+}$ and the dimers formed
 are actually compact---just like for large coupling in D=3.
 Hence, we can again neglect the overlap of the different dimers and attendant kinetic
 energy due to antisymmetrization in the spin up coordinates
$\{{\bf r}_i\}$ and separately in the spin down positions $\{{\bf r}'_j\}$.
 The strong coupling result Eq. (\ref{dimers}) applies and we have:
 $ E/N = - \frac{|B \cdot E|}{2}  \le 0 $. Thus, 
we find that $\xi$ may vanish (or become negative) 
for all dimensions greater than four,
\begin{equation}
  \xi \le 0 \; , \;\; D \ge 4.
\label{4D} 
\end{equation}
In the marginal dimension $D=4$,
the size of the dimers is just barely
$r_{0}$ (where the attractive potentials
operate) and the upper bound may be more easily saturated
with a potentially vanishing $\xi(D=4)$.

\subsubsection{A Variational Upper Bound} 

To substantiate the above arguments, and directly derive 
Eq.(\ref{4D}), we next employ a 2n-body ansatz wave
function $\Psi({\bf r}_1,{\bf r}_2,{\bf r}_i, \ldots {\bf r}_n;
{\bf r}'_1, \ldots ,{\bf r}'_j, \ldots {\bf r}'_n)$ and show
that in all continuous dimensions $D > 4$, we have $H|\Psi \rangle=0$. 
The ansatz $\Psi$ has the form of a
Slater ``orbital" determinant. It represents the state of n
dimers, say, $\overline{\phi}({\bf r}_1-{\bf r}'_1) \ldots \overline{\phi}({\bf
r}_n-{\bf r}'_n)$ with $\overline{\phi}(|{\bf r}_i-{\bf r}'_j|)$ the zero
energy bound state, antisymmetrized over all the n! permutations:
$ P=i \rightarrow (p(i)), \; i=1, \ldots n$ of the n spin-up
atoms: \setcounter{equation}{7}
\begin{equation}
  \Psi=\sum_{P} (-1)^P \Phi(P)\, ; \;\;\;\; \Phi(P) =
 \prod_{i=1...n} \overline{\phi}({\bf r}_i-{\bf r}'_{p(i)})
\label{PsiSigmaEq8}
\end{equation}
with $(-1)^P$ the parity of the permutation $P$.
Here, $\overline{\phi}({\bf r}_i-{\bf r}'_{p(i)})$ 
represent normalized, zero energy, pair wavefunctions.

The further required antisymmetrization over the n spin-down atoms
 simply scales $\Psi$: ~ $\Psi \to [(n!) \Psi]$ and is redundant. 
We note that $\Psi$ can be expressed as
a determinant,
\begin{eqnarray}
\Psi(\bf{r}_{1}, ...., \bf{r}_{n}; \bf{r'}_{1}, ..., \bf{r'}_{n}) 
= \nonumber
\\  \det \left( \begin{array}{cccc}
\overline{\phi}(\bf{r}_{1} - \bf{r'}_{1}) & 
\overline{\phi}(\bf{r}_{1}- \bf{r'}_{2}) & ... & 
\overline{\phi}(\bf{r}_{1} - \bf{r'}_{n}) \\
\overline{\phi}(\bf{r}_{2}- \bf{r'}_{1}) 
& \overline{\phi}(\bf{r}_{2} - \bf{r'}_{2}) & ... &
\overline{\phi}(\bf{r}_{2} - \bf{r'}_{n}) \\
. & . & ... & . \\
\overline{\phi}(\bf{r}_{n} - \bf{r'}_{1}) & 
\overline{\phi}(\bf{r}_{n} - \bf{r'}_{2}) & ... &
\overline{\phi}(\bf{r'}_{n} - \bf{r'}_{n})
\end{array} \right).
\label{DetMijEq9}
\end{eqnarray}
We can use the variational ansatz 
$\Psi$ in any dimension D to obtain a variational upper
bound on the N=2n-body ground state energy and on the corresponding
$\xi$:
\begin{equation}
  E(2n) = \Big[ \frac{2n D}{D+2}  \varepsilon_F  \xi \ \Big] \le \, 
\frac{\langle \Psi|H|\Psi \rangle}{\langle \Psi|\Psi \rangle}.
\label{E2n2nEq10}
\end{equation}
Clearly,
\begin{equation}
  \langle \Psi|H|\Psi \rangle = \sum_{P,P'} \langle  \Phi(P)|
H|\Phi(P') \rangle,
\label{PsiHPsiEq11}
\end{equation}
with a similar expression for $\langle \Psi|\Psi \rangle$.

The local potential limit, say, $r_0 \rightarrow 0$ in the
``square well'' spherical potential
\begin{equation}
 V({\bf r})= - V_0 \Theta(r_{0}-|{\bf r}|)
\label{VrV0Eq12}
\end{equation}
implies for all dimensions $D>2$ the following ``selection rule"
for non-vanishing $\langle \Phi(P')|V_{i,j}|\Phi(P) \rangle$
matrix elements appearing when the 2n-body Hamiltonian (1) is
substituted in Eq. (\ref{PsiHPsiEq11}) above.

Before going into the details of the variational 
calculations, let us quickly give the reader a 
glimpse of the final results derived 
below and their implication.
The calculation detailed in this
subsection illustrates that 
for the value of $V_{0}$ in Eq.(\ref{VrV0Eq12})
(the analog of $g$ in Eq.(\ref{Hamiltonian}))
which secures a single zero-energy bound,
state we have a variational state (i.e. $| \Psi \rangle$
of Eq.(\ref{DetMijEq9})) for which the corresponding 
energy $E_{var}=0$.
This result then implies (by Eqs.(\ref{trivDim},\ref{E2n2nEq10}))
that $\xi \le 0$.

For any $V_{i,j}$ among the $n^2$ spin-up--spin-down
potentials we have, for all $D>2$ with a local $V({\bf r})$ producing
one zero energy bound state, the following relation
\begin{equation}
 \langle \Phi(P')|V_{i,j}|\Phi(P) \rangle =0 {\rm \; if \; no \;} p(i)=j
 {\rm \; \underline{and} \; no\;}  p'(i)=j.
\label{PhiP'VijEq13}
\end{equation}
To prove this claim, we perform first (among the 2n $\; d^D \, {\bf r}_k
d^D \, {\bf r}'_l $ integrations involved in evaluating the above
matrix element) the integrals $\int d^D \, {\bf r}_i \int d^D
\, {\bf r}'_j$ over the coordinates ${\bf r}_i$ and ${\bf r}'_j$
appearing in the particular potential term 
$V({\bf r}_i-{\bf r}'_j)$ considered.

Both ${\bf r}_i$ and ${\bf r}'_j$ appear in the product $\Phi(P)$
of Eq.(\ref{PsiSigmaEq8}) but, by assumption, in different
$\phi$ factors. The same holds for $\Phi(P')$. Thus, the
integration over ${\bf r}_i$ and ${\bf r}'_j$ is of the form,
\begin{eqnarray}
\int d^3{\bf r}_i \int d^3{\bf r}'_j 
[\overline{\phi}({\bf r}_i-{\bf r'}_{p'(i)}) 
\overline{\phi}({\bf r}_{p'^{-1}(j)}-{\bf r}'_j) \nonumber
\\ \times V({\bf r}_i-{\bf
r}'_j)  
\overline{\phi}({\bf r}_i-{\bf r'}_{p(i)})
\overline{\phi}({\bf
r}_{p^{-1}(j)}-{\bf r'}_j)], \label{integralsEq14}
\end{eqnarray}
with $p^{-1}$ the permutation inverse to p.
The condition for having one zero energy S wave bound state in the
square well potential of (\ref{VrV0Eq12}) in the D=3 case is
well known (see Fig.(\ref{Fig1bound})),
\begin{equation}
V_0= \hbar^2 \frac{(\pi^2/4)}{mr_0^2} \;. \label{VOEq15}
\end{equation}
By dimensional considerations, the same condition for one zero
energy bound state holds in all dimensions $D>2$ wherein the constant 
$\pi^2/4$ is replaced by some other, dimension dependent, 
numerical factor $c_D$,
\begin{equation}
V_0 = \hbar^2 \frac{c_D}{mr_0^2}. \label{VOcd}
\end{equation}
In the general $D$ dimensional problem, 
$c_{D}$ of Eq.(\ref{VOcd}) is fixed by normalization of 
the zero-energy wavefunction $\phi({\bf r})$.
In the appendix, we briefly present the
solution to the a zero-energy bound state
in a $D$-dimensional spherical potential
well.

By changing variables $({\bf r}_i,{\bf r}'_j) \rightarrow ({\bf r}_i,
{\bf r}_{i,j})$ with ${\bf r}_{i,j} = ({\bf r}_i-{\bf r}'_j)$,
the local $V({\bf r}_{i,j})$ implies that ${\bf r}_i={\bf r}'_j$
in the arguments of all the four $\phi$ factors appearing above
which generically keep finite arguments and bound $\phi$ values.
Neglecting small variations of $\phi({\bf r})$ away from ${\bf r}
\sim 0$, the integral $\int d^3{\bf r} \, V({\bf r})$ over the D
dimensional sphere of radius $r_0$ where the square well
potential is non-vanishing yields the factor
 $V_0  r_0^D \Omega_{D}$. This multiplicative factor 
is, by virtue of Eq.(\ref{VOcd}), proportional to $[c_D r_0^{(D-2)}]$. 
In dimensions $D>2$, this factor vanishes as $r_{0} \to 0$ 
and, as claimed earlier, so does the complete matrix
element. This is {\it not} the case in dimensions $D<2$ for which 
$\xi=1$.

Next, we consider dimensions $D>4$. In this case, the zero energy
bound states 
\begin{eqnarray}
\overline{\phi}({\bf r}_{i}, {\bf r'}_{p(i)}) 
= {L^{-D/2}} \phi({\bf r}_{i} - {\bf r'}_{p(i)}),
\end{eqnarray} with 
(see the Appendix for details),
\begin{eqnarray}
\phi({\bf r}_i-{\bf r}'_{p(i)}) = A_{>}|{\bf r}_i-{\bf
r}'_{p(i)}|^{-(D-2)}  \nonumber
\\  {\mbox{for}} \; |{\bf{r}}| \equiv 
|{\bf r_{i}- r`_{p(i)}}| > 
r_0 ;\nonumber
\\ {\mbox{and}} \;\; \phi({\bf r}_i-{\bf r}'_{p(i)})
                  \sim A_{>} (r_0^{-(D-2)})  \nonumber
\\ {\rm for} \;
                  0<|{\bf r}| < r_0,
\label{D>4EQ16}
\end{eqnarray}
where $A_{>} \approx (r_0^{(D/2)-2})$. Exact forms
are provided by Eqs.(\ref{bigr},\ref{smallr}, 
\ref{BesselV}, \ref{amp_ratios}). The comparison
between Eq.(\ref{D>4EQ16}) and the exact forms
is provided in Eq.(\ref{fz}). 
The scaling of the results with $r_{0}$ (which we will
shortly obtain) becomes more transparent
with the use of Eq.(\ref{D>4EQ16}). With the 
incorporation of Eq.(\ref{fz}) and its ensuing discussion,
the results which we will obtain using Eq.(\ref{D>4EQ16}) 
will further enable
as a rigorous upper bound on the matrix value elements to be discussed. 
As seen from Eq.(\ref{D>4EQ16}), the wavefunctions $\phi$ are 
strongly localized at $|{\bf r}| \sim r_0$  implying stronger
selection rules for non-zero matrix elements:
\begin{equation}
 \langle \Phi(P')|\Phi(P) \rangle  =  0 \;\; {\rm  unless} \; P=P',
\label{PhiPhi17a}
\end{equation}
\vspace*{-.15in}
and 
\begin{equation}
 \langle \Phi(P')|V(r_i-r'_j)|\Phi(P) \rangle  =  0 \;\; {\rm unless} \;
j=p(i)=p'(i).  \label{PhiVPhi17b}
\end{equation}

Eq.(\ref{PhiVPhi17b}) implies that also $\langle \Psi|H|\Psi\rangle$, 
the numerator  of Eq.(\ref{E2n2nEq10}), has only diagonal
contributions. Namely,
\begin{equation}
  \langle \Phi(P')|H|\Phi(P) \rangle = 0 \;\; {\rm unless} \; P=P' \, .
\label{P=P'}
\end{equation}

To illustrate Eq.(\ref{PhiPhi17a}), let us assume that $P$ and $P'$ differ
minimally: $p(i)=p'(i)$ for all  $i>2$, but $p(1)=1$,  $p(2)=2$
and $p'(1)=2$, $p'(2)=1$. The overlap in Eq.(\ref{PhiPhi17a}) is
then
\begin{eqnarray}
 L^{-2D} \int d^D{\bf r}_1  d^D{\bf r}'_1  d^D{\bf r}_2
   d^D{\bf r}'_2 
\phi({\bf r}_1-{\bf r}'_1)  \phi({\bf r}_2-{\bf r}'_2) \nonumber
\\ \times 
 \phi({\bf r}_1-{\bf r}'_2) \phi({\bf r}_2-{\bf r}'_1) \; \nonumber
\\ L^{-nD} \prod_{i,j>2} \int d^{D}{\bf r}_{i} d^{D}{\bf r'}_{j}
\langle \phi(i,p(i))|\phi(i,p(i)) \rangle. \label{n-2ProdEq.18}
\end{eqnarray}

The last line of Eq.(\ref{n-2ProdEq.18}) denotes a normalized 
integral (equal to one). In the first integral over the 4 variables
$\{{\bf r}_{1,2}, {\bf r'}_{1,2}\}$, the number of integration
variables is equal to the number of wavefunctions 
$\{\phi({\bf r}_{i} - {\bf r'}_{j})\}$ appearing in the 
integrand. For each of the two particle 
wavefunctions $\phi({\bf r}_{i} - {\bf r'}_{j})$, we insert
Eq.(\ref{D>4EQ16}) (or the exact Eqs.(\ref{bigr},\ref{smallr}) 
derived within the appendix) with $A_{>} \approx r_{0}^{(D/2)-2}$. 
Insofar as scaling with $r_{0}$ is concerned, 
all integrals over products of wavefunction forms
of Eq.(\ref{D>4EQ16}) multiplying $A_{>}$ amount to
product of either integrals of the type $I_{>} = \int_{|{\bf r}|>r_{0}} 
d^{D} {\bf r} |r|^{2-D}$ or of the form $I_{<} = r_{0}^{2-D}
\int_{|{\bf r}| <r_{0}} d^{D} {\bf r} $.
Constant factors aside, both $I_{>}$ and $I_{<}$ scale as $r_{0}^{2}$.
The four normalization prefactors of $A_{>}$
(with $A_{>} \approx r_{0}^{(D/2)-2}$),
originating from the four factors of $\phi$ in the top two lines of 
Eq.(\ref{n-2ProdEq.18}),
lead to an additional factor of $r_{0}^{2(D-4)}$ in tow. Thus,
in the final analysis, when the integrals are segregated into 
all possible terms for $|{\bf r}_{i} - {\bf r'}_{j}| 
> r_{0}$ and for $|{\bf r}_{i} - {\bf r'}_{j}| < r_{0}$ and 
Eq.(\ref{D>4EQ16}) or Eqs.(\ref{bigr}, \ref{smallr})) are
inserted, we find that all terms scale as ${\cal A} \approx r_{0}^{2D}
\to 0$ as $r_{0} \to 0$.
This signals that the factor in the first two lines 
of Eq.(\ref{n-2ProdEq.18}) vanishes as $r_{0} \to 0$
and thus so does the entire overlap of Eq.(\ref{PhiPhi17a}).
We can readily verify that for larger mismatches between $P$
and $P'$ the overlap vanishes as a higher power of $r_0$.
Now, in the  diagonal case, with $p'(1) =1$ and $p'(2)=2$ 
the product of the four normalization
factor cancels and $\langle \Phi(P)| \Phi(P) \rangle =1$. 
\cite{explainnormal} 

Turning to Eq.(\ref{PhiVPhi17b}), we assume that $p(i)=j$ but $p'(i)
\neq j$ and prove that the matrix element $\langle P'|V_{i,j}|P \rangle$ 
vanishes for $D>4$. Recall that when also $p(i) \neq j$ we showed
that this matrix element vanishes in {\it all} $D >2$. Unlike
that previous case, we have here only three (rather than four)
$\phi$ functions in which ${\bf r}_i$ and/or ${\bf r}'_j$ appear.
Factors of $L$ aside, the 
relevant two integrations on ${\bf r}_i$ and ${\bf r}'_j$ in
the analog of Eq.(\ref{integralsEq14}) are now:
\begin{widetext}
\begin{eqnarray}
 & & \int d^D{\bf r}_i \int d^D{\bf r}'_j ~ V({\bf r}_i-{\bf r}'_j) 
\phi({\bf r}_i-{\bf r}'_{p(i)})\phi({\bf r}_{p'^{-1}(j)}-{\bf r}'_j)
 \phi({\bf r}_i-{\bf r}'_j) \nonumber \\
 & & = A_{<} V_0 \int d^D{\bf r_i} \int d^D{\bf r}
 ~ \Theta(r_0-|{\bf r}|)
[(\kappa r)^{1-\frac{D}{2}} J_{\frac{D}{2}-1}(\kappa r)] 
\phi({\bf r}_i-{\bf r}'_{p(i)}) \phi({\bf r}_{p'^{-1}(j)} - {\bf r}'_j),
\label{rel2integrationsEq.19}
\end{eqnarray}
\end{widetext}
where we invoke Eq.(\ref{smallr}). In particular, as
seen from the appendix, the wavenumber 
$\kappa = c_{D}^{1/2}/r_{0}$. 
In Eq.(\ref{rel2integrationsEq.19}), we employ, once again, 
$p^{-1}$ for the inverse permutation, ${\bf r}
= {\bf r}_i-{\bf r}'_j $ is the argument of the square well
potential, and $\phi({\bf r}_i-{\bf r}'_j)=\phi({\bf r})$. 
In what briefly follows, to flesh out more crisply the simple scaling
form of this term, we may replace the exact Bessel function form
of Eq.(\ref{smallr}) and instead invoke Eq.(\ref{D>4EQ16}) 
to replace the last factor $(\phi({\bf r}))$ for $r<r_{0}$ 
in the upper line of Eq.(\ref{rel2integrationsEq.19}).
The result which we will obtain in
this fashion will go unchanged if we employ the exact Bessel function 
form of Eq.(\ref{rel2integrationsEq.19}), see \cite{explainproof}. 
With the insertion of Eq.(\ref{D>4EQ16}), the $d^D{\bf r}$ 
integration of Eq.(\ref{rel2integrationsEq.19})
yields
\begin{eqnarray}
  V_0 \int_0^{r_0} d^D{\bf r} ~ [ r_0^{\frac{D}{2}-2}
   r_{0}^{(2-D)}]  = \Omega_{D} V_0 \, r_0^2
r_0^{\frac{D}{2}-2}\nonumber
\\ = \frac{\hbar^{2}}{m} \Omega_{D} c_D \, r_0^{\frac{D-4}{2}} \to 0   \;\;\;\;
(\mbox{for $D>4$}) \; , \label{VoIntEq.20}
\end{eqnarray}
in the limit $r_{0} \to 0$. 
Here, the definition of $c_{D}$ was 
invoked from Eq.(\ref{VOcd}). The peculiar form given in
the square braces of Eq.(\ref{VoIntEq.20})
follows from Eq.(\ref{D>4EQ16}) with the approximate
$A_{>}\approx (r_0^{(D/2)-2})$. The incorporation of the exact Bessel 
function form of Eq.(\ref{smallr}) in Eq.(\ref{rel2integrationsEq.19}) 
whose derivation is detailed in the appendix does not alter this simple 
result \cite{explainproof}. This concludes the proof of Eq.(\ref{PhiVPhi17b}).

With all shown thus far, vanishing in Eq.(\ref{rel2integrationsEq.19})
may be avoided only when $p'(i)=j=p(i)$. In this case, the
${\bf r}_i$ and ${\bf r}'_j$ integrations involve only two identical
$\phi$ functions and become (invoking Eq.(\ref{smallr}) once again),
\begin{eqnarray}
  \int d^D{\bf r}_i d^D{\bf r}~ \phi^{2}({\bf r}) \, V_0
    \cdot \Theta({\bf r}_0-|{\bf r}|) \nonumber
\\= V_0 A_{<}^{2} \int d^D{\bf r}_i \int_{0<|{\bf r}|
  <r_0} d^D{\bf r}~~
[(\kappa r)^{1-\frac{D}{2}} J_{\frac{D}{2}-1}(\kappa r)]^{2}  \nonumber
\\ \sim \int d^D{\bf r}_i \, V_0. \label{identicalphiEq.21}
\end{eqnarray}

By virtue of 
the zero-energy Schr\"odinger equation that $\phi({\bf r})$
satisfy, the contribution of Eq.(\ref{identicalphiEq.21})
(albeit divergent) 
identically cancels against the kinetic terms  
$T_i+T'_j = - (\hbar^2/2m) ({\bf \nabla}^2_i +
{\bf \nabla}'^2_j)$. The coordinates 
${\bf r}_i$ and ${\bf r}'_j$ now appear 
in both $\Phi(P)$ and $\Phi'(P)$ only in the above two
identical $\phi$ functions on the left and on the right, $T_i +
T'_{j}$ can be paired with the $V_{i,j}$ considered here into
$H_{i,j}=T_i+T'_j+V_{i,j}$. In the above shorthand notation, 
the matrix elements of $H_{i,j}$
are
\begin{widetext}
\begin{equation}
 \langle \Phi(P')|H_{i,j}|\Phi(P) \rangle =  \langle \phi(1,p'(1)) \ldots
  \phi(n,p'(n))\;|\;V_{i,j}-\frac{\hbar^2}{2m}(\nabla^2_i 
+ \nabla'^2_j)\;|\;
 \phi(1,p(1)) \ldots \phi(n,p(n)) \rangle.
\label{HijshorthandEq22}
\end{equation}
\end{widetext}
The Schr\"odinger equation for $\phi$ with the assumed small
binding energy (eventually $\epsilon \to 0$) which defines our problem is:
\begin{equation}
  [\frac{-\hbar^2({\bf \nabla}^2_i + {\bf \nabla}'^2_j)}{2m} 
+ V({\bf r}_i-{\bf r}'_i)]
  \phi(|{\bf r}_i - {\bf r}'_j|) = -\epsilon\phi(|{\bf r}_i - {\bf r}'_j|),
\label{h2DeltaEq23}
\end{equation}
or $H_{i,j}\phi_{i,j} =-\epsilon \phi_{i,j}$. Consequently, the 
matrix element of $H_{i,j}$ between the two $\phi_{i,j}$
in $\Phi(P)$ and in $\Phi(P')$ is just a number ($\epsilon$).
Omitting, then, $\phi_{i,j}=\phi(i,p(i)) = \phi(i,p'(i))$ from
the product of n functions $\phi$ in $\Phi(P)$ and $\Phi(P')$ 
leaves an overlap of
two monomials each composed of a product of the remaining $(n-1)$
orbital wavefunctions $\phi$ in both the bra and in the ket, i.e.
$\langle \Phi(P'_{n-1})|\Phi(P_{n-1}) \rangle$ with $P_{n-1}$
and $P'_{n-1}$, the
permutations of the remaining $(n-1)$ elements obtained by removing
$i \rightarrow p(i)=j$  and $i \rightarrow p'(i)=j $. Eq.(\ref{PhiPhi17a}) 
implies that this overlap vanishes unless these
remaining permutations are identical. Inserting 
$i \rightarrow p(i)=j$ and $i \rightarrow p'(i)=j$ implies that
$P=P'$. It follows that in the numerator of 
Eq.(\ref{E2n2nEq10}) we have,
similar to the denominator, only diagonal terms.

By definition, the permutations $(i,p(i))$ reproduce the $n$
distinct $(i,j)$ terms for given $i$. We may, therefore,
operate on $|\Phi(P) \rangle$ on the right  with the $n$  
potential terms $V(i,j)$ and the
$n$ kinetic energy operators 
$T_{i,j}$ with matching $\phi(i,p(i))$ terms in $\Phi(P)$. In
the process we employ each of the Laplacian terms 
($\{\nabla^2_i\}$  and $\{\nabla'^2_j\}$)
exactly once. All further unmatched $V(i,j)$ have vanishing matrix
elements.

Hence,
\begin{equation}
 H|\Phi(P) \rangle = -n\epsilon |\Phi(P)\rangle.
\label{PHPhiPEq24}
\end{equation}
This holds for each $\Phi(P)$ and, hence, for the full Slater
orbital ansatz of Eq.(\ref{DetMijEq9}) $\Psi$ in $D>4$ dimensions,
\begin{equation}
 H|\Psi \rangle = E(2n) = -n\epsilon|\Psi\rangle  \, .
\label{HPsiEq25}
\end{equation}

Thus, the total energy of the system is just that of 
$n$ independent Bose-Einstein condensed
dimers. When the binding energy 
$\epsilon \rightarrow 0$ we have $E(2n) \rightarrow 0
$ and for the variational orbital Slater determinant
of Eq.(\ref{DetMijEq9}), we find  $\xi_{var}(D>4) = 0$ as claimed. 
By the standard variational theorem, this proves that that for continuous
$D$, 
\begin{eqnarray}
\xi(D>4) \le 0. 
\end{eqnarray}
The exact variational
result fortifies the conclusion arrived
at earlier (subsection(\ref{norm}), with Eq.(\ref{4D}) in particular) 
via normalization considerations. In the final analysis, 
normalization was present in our slightly elaborate
variational bound and ultimately 
dictated the scaling form of our expressions
with $r_{0}$ (see, e.g., 
Eq.(\ref{n-2ProdEq.18}), \cite{explainnormal})
as $r_{0} \to 0$.

\section{Heuristic considerations for \boldmath{$\xi(D=3)$}}  
\label{threeD}

The extension of the problem to arbitrary continuous 
dimension $D$ allowed us to firmly establish 
that $\xi =1$ in $D \le 2$ and $\xi =0$ in $D \ge 4$ (with the 
particular point $D=4$ accessed by general normalization conditions;
our rigorous variational bound $\xi(D>4) \le 0$ held for all continuous $D>4$).
These results suggest that in a lowest
order $\epsilon$ expansion interpolating between these
two limits, $\xi(D=3) \approx 1/2$. [Or, if the upper bound 
in the variational inequality for all continuous large $D$,
$\xi(D>4) \le 0$ is not saturated for the 
marginal $D=4$,
that, more generally, $\xi(D=3) \le 1/2$.] This is, remarkably,
not far from the results obtained
by very intensive numerical work \cite{CCPS}
(which led to an upper bound of $\xi(D=3) \approx 0.44$). 
Unfortunately, we have not been able to
attain direct potent bounds on the three dimensional
problem which are close to the numerically
attained values. In the current section, we
attack the harder $D=3$
problem by a heuristic approach.
This approach, although imprecise, may shed light on
some of the observed physics in three dimensions.
Similar to the simple minded dimensional interpolation
result it, too, suggests that $\xi(D=3) \approx 1/2$. As an
additional bonus, this 
allows for a heuristic argument for
even-odd oscillations numerically observed.

Unlike the normalization of the zero energy bound state, which for
$|{\bf r}| > r_0$ is uniform
 in $|{\bf r}|=|{\bf r}_i-{\bf r}'_j|$, the kinetic energy,
\begin{eqnarray} 
\int d^{3}{\bf r}(\nabla \phi({\bf r}))^2 =
 \int d^{3}{\bf r} \frac{A_{>}^2}{|{\bf r}|^4},
\label{3Dint+}
\end{eqnarray}
is mostly concentrated at small $|{\bf r}| \sim r_0 <<d$
values. In the last equality of Eq.(\ref{3Dint+}), 
we inserted the well known ($D=3$) zero-energy S-wave wavefunction
form outside a potential well, $\phi = \frac{A_{>}}{|{\bf r}|}$. 
As a consequence of this concentration of kinetic energy, 
a zero energy bound state of a mixed $(i,j')$ pair,
can manifest over any smooth background function $B$ 
of all other variables by a $\frac{1}{|{\bf r}_i-{\bf r}'_j|}$
 enhancement at distances $|{\bf r}_i-{\bf r}'_j|$ smaller 
than typical particle spacing.
 (Due to antisymmetry, the i$^{th}$ spin up 
atom is equally likely to bind with any one of the
 n down spin atoms at ${\bf r}'_j$, with $j =1, ..., n$.)

 This is indeed the case for the wave functions 
serving as the staring point of the above MC
 where correlations in mixed pairs are introduced via a Jastrow product $J$,
\begin{equation}
\Psi = B \times J \equiv B \prod_{j=1,..n} \,  \phi(|{\bf{r}}_{i}-{\bf{r}}'_j|),
\label {J}   
\end{equation}
where the background B of free Fermion Slater determinants or BCS
form accounts for the antisymmetrization.

 The other $(n-1)$ factors of the form 
$ \; $   $ \frac{1}{|{\bf r}_i-{\bf r}'_k|}$
with $k$ different from $j$
 still allow for
 $\Psi \sim \frac{C}{|{\bf r}_i-{\bf r}'_j|}$
for  $|{\bf r}_i-{\bf r}'_j| <<d$. Thus, let the remaining ($n-1$)
 down spin atoms
 form a cubic lattice of spacing $d$, the triply periodic product,
 $\prod_{j=1...n-1} \, \phi(|{\bf r}'_j-{\bf r}_i|)$, is then invariant under the
point symmetry group of the simple cubic lattice. Its finite
extrema are at the centers of the n cubes  (each of volume 
$d^3$) and C varies
slowly for $|{\bf r}|<<d$.

 This suggests (yet does {\it not} justify) the following ad-hoc 
approach which yields $\xi=1/2$. The 
potentials $V(|{\bf r}_i-{\bf r}'_j|)$ involve atom pairs yet the kinetic energy is a sum over
 single atoms:
\begin{equation}
  \sum_{I=1,..N=2n}\, \frac{\bf{p}^{2}_I}{2m} \,=\,
\sum_{i=1,...n}\, \frac{\bf{p}_i^2}{2m}\,
 + \,\sum_{j=1,...n} \,\frac{(\bf{p}'_j)^2}{2m}.
\label{KE} 
\end{equation}
  We formally associate kinetic energies with pairs by using the identity :
\begin{equation}
 \sum \, \frac{{\bf{p}}_I^2}{2m} \, = \, \frac{1}{2m (N+1)} \, \sum_{I<J=1,2...N}
  \, ({\bf{p}}_I- {\bf{p}}_J)^2,
\label{KEvarpairs}  
\end{equation}
 where ${\bf{P}}_{tot}^2 =(\sum {\bf{p}}_I)^2=0$ was subtracted cancelling
 all $I<J$ terms $({\bf p}_I \cdot {\bf p}_J)$.
 Next, we separate the contribution to the Hamiltonian of same (up-up, down-down) and
 mixed spin pairs with the potentials included in the second term:
\begin{eqnarray}
 H & = & H_{same}  + \, H_{mixed}  \nonumber
\\ & = &   \frac{1}{2m (N+1)}  \sum_{i<l=1,...n}
  ({\bf{p}}_i-{\bf{p}}_l)^2  \nonumber
\\ & \;\;\;\;\;\;\;\;+ & \frac{1}{2m (N+1)}
\sum_{j<k=1,...n}  ({\bf{p}}'_j-{\bf{p}}'_k)^2  \nonumber \\
  & \;\;\;\;\;\;\;\; + &  \sum_{i,j=1,...n} \, [\frac{({\bf{p}}_i-{\bf{p}}'_j)^2}{2m(N+1)}  -
 gV({\bf r}_i,{\bf r}'_j)].
\label{updownmixpairs}  
\end{eqnarray}

 For $g=0$ the kinetic terms in $H_{same}$ and $H_{mixed}$ are equal up to 1/N corrections.

Next, we note that if the following assumptions are made:

 (i) Just as for isolated up down pairs also in the N-body ground state
 $\Psi$,
the attractions between different atoms cancel the kinetic energy
of the relative motion, i.e.,  $\langle \Psi |H_{mixed}|\Psi
\rangle = 0$, and that

 (ii)  turning on the coupling $g$ does not change the expectation value of
$H_{same}$, then:

\begin{eqnarray}
  (E/N)|_{N=even} = \frac{(n-1)}{(2n-1)} \; \frac{3}{5} \; \varepsilon_F  \nonumber
\\ \simeq \frac{1}{2}[1 + \frac{1}{N}]\; \frac{3}{5} \; \varepsilon_F.
\label{evenN} 
\end{eqnarray}
  For odd $N$, say $n+1$ spin up and $n$ spin down atoms, only $n=N/2-1/2$
bound pairs can form
 The decrease of potential attraction energy by $(N-1)/N$  produces a ``gap"
\begin{equation}
                 \frac{E}{N}|_{N=2n+1}-\frac{E}{N}|_{N=2n} \approx 
\varepsilon_F,
\label{oddN}   
\end{equation}
 consistent with the finding of the MC calculations. Furthermore,
\begin{equation}
   \xi \to 1/2 \; \;\; (\mbox{as~~} N\rightarrow{\infty}), 
\end{equation} 
which is barely consistent with the variational bound $\xi< 0.43-0.45$ from
the above calculations.\cite{Cohen}

\section{Conclusions} 
In conclusion, we examined an extension 
of the BCS to BEC crossover problem to arbitrary
dimension $D$. We report new results on the ground
state energy per particle at the crossover point by employing
direct variational bounds in high dimensions
and by considering the consequences 
of localization in low dimensions. In particular, we
find that the ground
state energy per particle at the onset of
the BCS to BEC crossover is zero (or negative) 
in all dimensions $D \ge  4$ while it is the energy of a
free Fermi system in 
all dimensions $D \le 2$. The interpolation of these
bounds to the physical 
three dimensional problem leads to an 
energy per particle which is half that of 
the free fermion energy and is close to current
numerical results. We outlined
a simple heuristic argument
for the same result which further mandates
even-odd variations which are indeed
seen numerically.

{\bf{Acknowledgments}}

  We benefited from discussions with G. A. Baker,
A. Casher, R. Furnstahl, A. Schwenk and would like to thank J.
Carlson, G. Ortiz and E. M. Timmermans, the organizers of the 2004
Los Alamos/Santa Fe workshop on cold atoms.

\appendix

\section{The Zero Energy Bound State Problem in a D-dimensional
spherical potential well}
\label{ap1}

Here, we provide the solution of the zero energy 
bound state problem in $D$ dimensions for the spherical 
potential well of Eq.(\ref{VrV0Eq12}) wherein the 
depth of the well ($V_{0}$) is tuned to get a zero-energy bound
state. The solution is a straightforward extension of the 
standard $D=1,2,3$ dimensional spherical potential well problems.
In these, 
without any angular dependence, in the 
the ``S-wave'' representation (i.e. the scalar (``$l=0$'') representation
of the $SO(D)$ rotation group), the D-dimensional Laplacian is 
$\nabla^{2} = [ \frac{d^{2}}{dr^{2}} + \frac{D-1}{r} \frac{d}{dr}]$.
This appendix explicitly illustrates how the numerical constant 
$c_{D}$ of Eq.(\ref{VOcd}) may be determined (from the implicit
Eq.(\ref{BesselV})). Its results further allow us to compare the approximate
form of Eq.(\ref{D>4EQ16}) introduced within the text to allow a clear
understanding of the scaling form of the overlap integrals with 
the exact form of $\phi(r)$ (Eqs.(\ref{bigr},\ref{smallr})).
In the aftermath, it will be shown that the incorporation
of the exact functional from does not change the scaling results
derived in the main text (following Eq.(\ref{D>4EQ16})).

We proceed with the solution of the translationally 
invariant problem specified
by the spherical potential well 
of Eq.(\ref{VrV0Eq12}). By translational invariance
of the center of mass, the two body wavefunction
\begin{eqnarray}
\overline{\phi}({\bf x}, {\bf x}')
= L^{-D/2} \phi({\bf r})
\end{eqnarray} 
with ${\bf r} \equiv  {\bf x}' - {\bf x}$. In the 
potential-free region $(r=|{\bf r}|>r_{0})$, the wavefunction 
$\phi$ satisfies a single particle Schr\"odinger equation 
with a reduced mass $\mu = m/2$,
\begin{equation}
\frac{d^{2}}{dr^{2}} \phi(r) + 
\frac{D-1}{r} \frac{d}{dr} \phi(r) + \kappa^{2} \phi(r) = 0,
\label{schr}
\end{equation}
with $\kappa=0$. Within the spherical potential
well ($r < r_{0}$), we have Eq.(\ref{schr}) with
$\kappa^{2} = (m V_{0})/\hbar^{2}$. We immediately find
that
\begin{eqnarray}
\phi(r>r_{0}) = A_{>} r^{2-D},
\label{bigr}
\end{eqnarray}
while within the potential well
\begin{eqnarray}
\phi(r<r_{0}) = A_{<}[(\kappa r)^{1-\frac{D}{2}} 
J_{\frac{D}{2}-1}(\kappa r)],  
\label{smallr}
\end{eqnarray}
with $\kappa$ and the constant prefactors $A_{>}$ and $A_{<}$ 
determined by continuity at 
$r=r_{0}$ and global normalization. Here, $J_{\frac{D}{2}-1}$ is
a Bessel function of order $(\frac{D}{2}-1)$. (Inserting
$J_{\frac{1}{2}}(\kappa r) = [(\kappa r)^{-1/2} \sin \kappa r]$, 
the pertinent
three dimensional result, illustrated in Fig.(\ref{Fig1bound}),
is recovered.) Continuity at $r=r_{0}$
restrains 
\begin{eqnarray}
\phi(r<r_{0}) \approx A_{>} r_{0}^{2-D}.
\label{contA}
\end{eqnarray}
For small $r$ (i.e. $r \to 0$), employing the asymptotic form
of the Bessel functions, 
we find that
\begin{eqnarray}
\phi(r \to 0) = \frac{A_{<}}{\Gamma(\frac{D}{2})2^{\frac{D}{2}-1}}.
\end{eqnarray} 
Anywhere within the potential well ($r <r_{0}$), the wave function 
\begin{eqnarray}
\phi(r<r_{0}) =  f(\frac{r}{r_{0}}) A_{>} r_{0}^{2-D},
\label{fz}
\end{eqnarray} 
with 
$f(0\le z \le 1)$  a bounded function satisfying $f(1) =1$ and 
$f(0) = \frac{A_{<}}{A_{>}} \frac{r_{0}^{D-2}}{\Gamma(\frac{D}{2})
2^{\frac{D}{2}-1}}$. Within the text following
Eq.(\ref{D>4EQ16}), we set $f$ to be one
and we attained vanishing overlaps
in Eq.(\ref{VoIntEq.20}) in the
limit $r_{0} \to 0$. The incorporation
of the more explicit, yet {\em bounded}, $f(z)$ 
given by Eq.(\ref{smallr}) does
not alter this result. Formally, our
results derived in the text can be made
rigorous by deriving an upper bound which
differs from the results given in Eq.(\ref{VoIntEq.20})
by a factor of $|f|_{\max}$, the maximal value of $|f|$. 
As Eq.(\ref{VoIntEq.20}) vanishes by simple scaling,
the incorporation of a finite factor leads to a vanishing
upper bound, proving that Eq.(\ref{rel2integrationsEq.19}) vanishes
in the limit of small $r_{0}$. Although inconsequential for this
bound (all that matters is its bounded norm), 
the prefactor $f(r/r_{0})$ is everywhere positive 
as the zero energy ground state is nodeless 
and of uniform sign.

For completeness, we now outline the general solution.
 Matching the two functional forms of $\phi(r)$ and their 
derivatives at $r=r_{0}$ (and simplifying with the 
aid of the standard Bessel function 
recursion relations) leads to the implicit equation
\begin{eqnarray}
(1- \frac{D}{2}) J_{\frac{D}{2}-1}(\kappa r_{0}) \nonumber
\\  =  \frac{1}{2} (\kappa r_{0})
[J_{\frac{D}{2}}(\kappa r_{0}) + J_{\frac{D}{2}- 2}(\kappa r_{0})].
\label{BesselV}
\end{eqnarray}
The solution of Eq.(\ref{BesselV}) enables a determination
of $\kappa$ and thus of the spherical potential depth $V_{0}$ and
the constant $c_{D}$ in Eq.(\ref{VOcd}) 
which ensures a zero-energy ``S-wave''
state in D dimensions.
The amplitudes $A_{>}$ and $A_{<}$ are then determined
by 
\begin{eqnarray}
\frac{A_{<}}{A_{>}} = \frac{r_{0}^{2-D}}{(\kappa r_{0})^{1 - \frac{D}{2}} 
J_{\frac{D}{2}-1}(\kappa r_{0})},
\label{amp_ratios}
\end{eqnarray}
in conjunction with the global normalization of $\phi({\bf r})$. 
Normalization demands that, up to dimension dependent
numerical constants, $A_{>} \approx r_{0}^{D/2-2}$
and consequently $A_{<} \approx r_{0}^{-D/2}$.


\begin{thebibliography}{99}

\bibitem{gas95} M. H. Andersen, J. R. Ensher, M. R. Matthews, C. E. Wieman,
and E. A. Cornell, Science {\bf 269}, 198 (1995); K. B. Davis,M.-O. Mewes, 
M. R. Andrews, N. J. van Druten, D. S. Durfee, D. M. Kurn, 
and W. Ketterle, Phys. Rev. Lett. {\bf 75}, 3969 (1995);
C. C. Bradley, C. A. Sackett, J. J. Tollett and R. G. Hulet, Phys. Rev. Lett.
{\bf 75}, 1687 (1995); C. C. Bradley, 
C. A. Sackett, and R. G. Hulet, Phys. Rev. Lett.
{\bf 78}, 985 (1997)


\bibitem{RGJ} C. A. Regal, M. Greiner and D. S. Jin, Phys. Rev. Lett. {\bf
92}, 040403 (2004);
M. Zwierlein, et.al., Phys. Rev. Lett. {\bf 92}, 120403 (2004).

\bibitem{DH} R. B. Diener and T-L. Ho,  
Phys. Rev. Lett. {\bf 94}, 090402 (2005) 

\bibitem{epsilonexpansion} K. G. Wilson, Phys. Rev. Lett. {\bf 28}, 548 
(1972);  K. G. Wilson and M. E. Fisher, 
Phys. Rev. Lett. {\bf 28}, 240
(1972);  K. G. Wilson and J. Kogut, Phys. Rep. {\bf 12}, 75
(1974); J. Zinn-Justin, {\em Quantum Field Theory and Critical Phenomena},
Oxford University Press (1989)

\bibitem{ChallProb} G. F. Bertsch, ``Challenge Problems in Many-Body
Physics",
http//:www.phys.washington.edu/$\sim$mbx/george.html (1998).

\bibitem{CCPS} J. Carlson, S.-Y. Chang, V. R. Pandharipande, and K.
E. Schmidt, Phys. Rev. Lett. {\bf 91}, 050401 (2003).

\bibitem{Baker} G. A. Baker, Jr., Phys. Rev. C {\bf 60}, 054311
(1999).

\bibitem{scale} The hierarchy of scales, $a>>d>>r_0$, which is 
implicit here may not suffice to
guarantee the same universal expression as E/N could also depend on $d^2/(a 
r_0)$. 

\bibitem{LeggetEngelbrecht}  A. J. Legget, in {\it Modern Trends in the
Theory
of Condensed Matter}, edited by A. Pekalski and R.Przystawa
(Springer-Verlag, Berlin, 1980); J. R. Engelbrecht,
M. Randeria, and C. sa de Melo, Phys. Rev. B {\bf 55}, 15 153 (1997).



\bibitem{LL} Landau, Lifshitz, {\it Quantum Mechanics}
(Pergamon Press), third edition (1977)

\bibitem{LdeLAndRDS}  E. H. Lieb and M. de Liano, J.
Math. Phys.{\bf 19},860
 (1978); M. Randeria, J.-M. Duan, and L.-Y. Shieh,
Phys. Rev. Lett.
{\bf 62}, 981 (1989).

\bibitem{MV}  W. Metzner and D. Voldhardt, Phys. Rev.
Lett. {\bf  62},  324
(1989).

\bibitem{Steel} J.W.Steel, arXiv:nucl- th/0010066
(2000).

\bibitem{SKC}T. Schaefer, C-W Kao, and S. R. Cotanch, arXiv:nucl-th/0504088
(2005)


\bibitem{gang4} E. Abrahams, P. W. Anderson, D. C. Licciardo, 
and T. V. Ramakrishnan, Phys. Rev. Lett. {\bf 42}, 673 (1979)

\bibitem{explainnormal} In this case, we recover (by definition)
the normalization integral. For a single up/down spin 
pair (${\bf x}, {\bf x}')$
this (with the aid of Eqs.(\ref{bigr}, \ref{smallr})) 
is of the forms 
\begin{eqnarray}
{\cal I}_{>} = A_{>}^{2} L^{-D} \int d^{D}{\bf R} \int_{r>r_{0}} \frac{d^{D}
{\bf r}}{r^{2(D-2)}} \nonumber
\\ \mbox{ ~or~~~~}
{\cal I}_{<} = A_{<}^{2} L^{-D} \int d^{D} {\bf R}  \int_{r<r_{0}} d^{D}{\bf r}
 \frac{[J_{\frac{D}{2} -1}(\kappa r)]^{2}}{ (\kappa r)^{D-2}},
\label{calI}
\end{eqnarray}
with ${\bf R} = ({\bf x}+ {\bf x}')/2$ 
the center of mass, whose integration 
cancels identically against the volume factors
of $L^{D}$. Due to the higher power 
of $r$ in the denominators (vis a vis
the integrals in Eq.(\ref{n-2ProdEq.18})),
a lower power of $r_{0}$ is attained in ${\cal I}_{>},{\cal I}_{<}$
with respect to their permuted counterparts (see below) . 
In ${\cal I}_{>}$, the scaling is that of $A_{>}^{2} r_{0}^{4-D}$
which mandates the relation $A_{>} \approx r_{0}^{D/2-2}$ appearing
in the text (Eq.(\ref{D>4EQ16})). The lower power of $r_{0}$ appearing
here in the normalization integrals when no permutations are 
performed relative to the power of $r_{0}$ resulting from 
any integral appearing with a permutation (i.e. the scaling of of the integrals
${\cal I}_{>}$ of Eq.(\ref{calI}) relative to that of $I_{>}$ and $I_{<}$
defined following Eq.(\ref{n-2ProdEq.18})))
ensures that all integrals of the type of Eq.(\ref{n-2ProdEq.18}))
vanish in the $r_{0} \to 0$ limit. 
 
\bibitem{explainproof}
Indeed, this may be rigorously proven by replacing the positive function 
$f(z)$ defined via Eq.(\ref{fz}), (with the scaled radial coordinate 
$z \equiv r/r_{0}\sim \kappa r$), by its maximal value 
(denoted henceforth by $f_{\max}$) 
in the interval $[0,1]$ to attain an upper bound on
the integral of Eq.(\ref{rel2integrationsEq.19}). This upper
bound differs from the result shown in Eq.(\ref{VoIntEq.20})
by a factor $f_{\max}$. As $f_{max}$ is finite,
the vanishing of Eq.(\ref{rel2integrationsEq.19}) cannot be avoided
in a spherical well of small radial extent $(r_{0} \to 0)$.

\bibitem{Cohen} This feature motivated the  extension to
unequal numbers of spin-up and spin-down atoms leading to a new
exact and interesting result for the asymmetry energy, T. D. Cohen, 
Phys. Rev. Lett. {\bf 95}, 120403 (2005)
(arXiv:cond-mat/0505080)

\end{thebibliography}
\end{document}